\documentstyle[graphicx,psfrag,twoside]{article}

\catcode`\@=11
\long\def\@makefntext#1{
\protect\noindent \hbox to 3.2pt {\hskip-.9pt  
$^{{\eightrm\@thefnmark}}$\hfil}#1\hfill}               

\def\thefootnote{\fnsymbol{footnote}}
\def\@makefnmark{\hbox to 0pt{$^{\@thefnmark}$\hss}}    
        
\def\ps@myheadings{\let\@mkboth\@gobbletwo
\def\@oddhead{\hbox{}
\rightmark\hfil\eightrm\thepage}   
\def\@oddfoot{}\def\@evenhead{\eightrm\thepage\hfil
\leftmark\hbox{}}\def\@evenfoot{}
\def\sectionmark##1{}\def\subsectionmark##1{}}

\oddsidemargin=\evensidemargin
\renewcommand{\thefootnote}{\fnsymbol{footnote}}

\newcounter{sectionc}\newcounter{subsectionc}\newcounter{subsubsectionc}
\renewcommand{\section}[1] {\vspace{12pt}\addtocounter{sectionc}{1} 
\setcounter{subsectionc}{0}\setcounter{subsubsectionc}{0}\noindent 
        {\tenbf\thesectionc. #1}\par\vspace{5pt}}
\renewcommand{\subsection}[1] {\vspace{12pt}\addtocounter{subsectionc}{1} 
        \setcounter{subsubsectionc}{0}\noindent 
        {\bf\thesectionc.\thesubsectionc. {\kern1pt \bfit #1}}\par\vspace{5pt}}
\renewcommand{\subsubsection}[1] {\vspace{12pt}\addtocounter{subsubsectionc}{1}
        \noindent{\tenrm\thesectionc.\thesubsectionc.\thesubsubsectionc.
        {\kern1pt \tenit #1}}\par\vspace{5pt}}
\newcommand{\nonumsection}[1] {\vspace{12pt}\noindent{\tenbf #1}
        \par\vspace{5pt}}

\newcounter{appendixc}
\newcounter{subappendixc}[appendixc]
\newcounter{subsubappendixc}[subappendixc]
\renewcommand{\thesubappendixc}{\Alph{appendixc}.\arabic{subappendixc}}
\renewcommand{\thesubsubappendixc}
        {\Alph{appendixc}.\arabic{subappendixc}.\arabic{subsubappendixc}}

\renewcommand{\appendix}[1] {\vspace{12pt}
        \refstepcounter{appendixc}
        \setcounter{figure}{0}
        \setcounter{table}{0}
        \setcounter{lemma}{0}
        \setcounter{theorem}{0}
        \setcounter{corollary}{0}
        \setcounter{definition}{0}
        \setcounter{equation}{0}
        \renewcommand{\thefigure}{\Alph{appendixc}.\arabic{figure}}
        \renewcommand{\thetable}{\Alph{appendixc}.\arabic{table}}
        \renewcommand{\theappendixc}{\Alph{appendixc}}
        \renewcommand{\thelemma}{\Alph{appendixc}.\arabic{lemma}}
        \renewcommand{\thetheorem}{\Alph{appendixc}.\arabic{theorem}}
        \renewcommand{\thedefinition}{\Alph{appendixc}.\arabic{definition}}
        \renewcommand{\thecorollary}{\Alph{appendixc}.\arabic{corollary}}
        \renewcommand{\theequation}{\Alph{appendixc}.\arabic{equation}}
        \noindent{\tenbf Appendix \theappendixc #1}\par\vspace{5pt}}
\newcommand{\subappendix}[1] {\vspace{12pt}
        \refstepcounter{subappendixc}
        \noindent{\bf Appendix \thesubappendixc. {\kern1pt \bfit #1}}
        \par\vspace{5pt}}
\newcommand{\subsubappendix}[1] {\vspace{12pt}
        \refstepcounter{subsubappendixc}
        \noindent{\rm Appendix \thesubsubappendixc. {\kern1pt \tenit #1}}
        \par\vspace{5pt}}

\newcommand{\textlineskip}{\baselineskip=13pt}
\newcommand{\smalllineskip}{\baselineskip=10pt}

\def\eightcirc{
\begin{picture}(0,0)
\put(4.4,1.8){\circle{6.5}}
\end{picture}}
\def\eightcopyright{\eightcirc\kern2.7pt\hbox{\eightrm c}}

\def\abstracts#1#2#3{{
        \centering{\begin{minipage}{4.5in}\baselineskip=10pt\footnotesize
        \parindent=0pt #1\par
        \parindent=15pt #2\par
        \parindent=15pt #3\par
        \end{minipage}}\par}} 

\def\keywords#1{{
       \centering{\begin{minipage}{4.5in}\baselineskip=10pt\footnotesize
       {\footnotesize\it Keywords}\/: #1
        \end{minipage}}\par}}

\newcommand{\bibit}{\nineit}
\newcommand{\bibbf}{\ninebf}
\renewenvironment{thebibliography}[1]
        {\frenchspacing
         \ninerm\baselineskip=11pt
         \begin{list}{\arabic{enumi}.}
        {\usecounter{enumi}\setlength{\parsep}{0pt}     
         \setlength{\leftmargin 17pt}{\rightmargin 0pt}   
         \setlength{\itemsep}{0pt} \settowidth
        {\labelwidth}{#1.}\sloppy}}{\end{list}}

\newcounter{itemlistc}
\newcounter{romanlistc}
\newcounter{alphlistc}
\newcounter{arabiclistc}

\newcommand{\fcaption}[1]{
        \refstepcounter{figure}
        \setbox\@tempboxa = \hbox{\footnotesize Fig.~\thefigure. #1}
        \ifdim \wd\@tempboxa > 5in
           {\begin{center}
        \parbox{5in}{\footnotesize\smalllineskip Fig.~\thefigure. #1}
            \end{center}}
        \else
             {\begin{center}
             {\footnotesize Fig.~\thefigure. #1}
              \end{center}}
        \fi}

\newcommand{\tcaption}[1]{
        \refstepcounter{table}
        \setbox\@tempboxa = \hbox{\footnotesize Table~\thetable. #1}
        \ifdim \wd\@tempboxa > 5in
           {\begin{center}
         \parbox{5in}{\footnotesize\smalllineskip Table~\thetable. #1}
            \end{center}}
        \else
             {\begin{center}
             {\footnotesize Table~\thetable. #1}
              \end{center}}
        \fi}

\def\@citex[#1]#2{\if@filesw\immediate\write\@auxout
        {\string\citation{#2}}\fi
\def\@citea{}\@cite{\@for\@citeb:=#2\do
        {\@citea\def\@citea{,}\@ifundefined
        {b@\@citeb}{{\bf ?}\@warning
        {Citation `\@citeb' on page \thepage \space undefined}}
        {\csname b@\@citeb\endcsname}}}{#1}}

\newif\if@cghi
\def\cite{\@cghitrue\@ifnextchar [{\@tempswatrue
        \@citex}{\@tempswafalse\@citex[]}}
\def\citelow{\@cghifalse\@ifnextchar [{\@tempswatrue
        \@citex}{\@tempswafalse\@citex[]}}
\def\@cite#1#2{{$\null^{#1}$\if@tempswa\typeout
        {IJCGA warning: optional citation argument 
        ignored: `#2'} \fi}}

\def\pmb#1{\setbox0=\hbox{#1}
        \kern-.025em\copy0\kern-\wd0
        \kern.05em\copy0\kern-\wd0
        \kern-.025em\raise.0433em\box0}

\def\fnt#1#2{\footnotetext{\kern-.3em
        {$^{\mbox{\scriptsize #1}}$}{#2}}}

\def\fpage#1{\begingroup
\voffset=.3in
\thispagestyle{empty}\begin{table}[b]\centerline{\footnotesize #1}
        \end{table}\endgroup}

\def\runninghead#1#2{\pagestyle{myheadings}
\markboth{{\protect\footnotesize\it{\quad #1}}\hfill}
{\hfill{\protect\footnotesize\it{#2\quad}}}}
\font\tenbf=cmbx10
\font\tenit=cmti10 
\font\tenit=cmti10
\font\bfit=cmbxti10 at 10pt
\font\ninebf=cmbx9
\font\ninerm=cmr9
\font\nineit=cmti9

\font\eightrm=cmr8

\def\lsym{\raise-3pt\hbox{\vbox{\tabskip0pt\offinterlineskip
        \halign{\tabskip0pt plus 1em
        ##\tabskip0pt\cr
        $\,\,<\,\,$\cr
        $\,\,\sim\,\,$\cr}}}}
\def\rsym{\raise-3pt\hbox{\vbox{\tabskip0pt\offinterlineskip
     \halign{\tabskip0pt plus 1em
      ##\tabskip0pt\cr
      $\,\,>\,\,$\cr
      $\,\,\sim\,\,$\cr}}}}
\def\qed{\hbox{${\vcenter{\vbox{                        
        \hrule height 0.4pt\hbox{\vrule width 0.4pt height 6pt
        \kern5pt\vrule width 0.4pt}\hrule height 0.4pt}}}$}}
\def\theequation{\thesection.\arabic{equation}}         
\renewcommand{\thefootnote}{\fnsymbol{footnote}}        

\begin{document}

\runninghead{S. Hauswirth}
{Perfect Discretizations of Differential Operators}

\normalsize\textlineskip
\thispagestyle{empty}
\setcounter{page}{1}


\vspace*{0.88truein}

\fpage{1}
\centerline{\bf PERFECT DISCRETIZATIONS OF DIFFERENTIAL OPERATORS}
\vspace*{0.37truein}
\centerline{\footnotesize SIMON HAUSWIRTH} 
\vspace*{0.015truein}
\centerline{\footnotesize\it Institute for Theoretical Physics,
University of Bern,} 
\baselineskip=10pt
\centerline{\footnotesize\it Sidlerstrasse 5, 3012 Bern, Switzerland}
\baselineskip=10pt
\centerline{\footnotesize\it E-mail: simon.hauswirth@itp.unibe.ch}

\vspace*{0.225truein}

\vspace*{0.21truein}
\abstracts{In this paper we investigate an approach for the numerical solution of
differential equations which is based on the perfect discretization of
actions. Such perfect discretizations show up at the fixed points of
renormalization group 
transformations. This technique of integrating out the high momentum
degrees of freedom 
with a path integral has been mainly used in lattice field
theory, therefore our study of its application to PDE's explores new
possibilities.  We
calculate the perfect discretized Laplace operator for several 
non-trivial boundary conditions analytically and numerically. Then we
construct a parametrization of the perfect Laplace operator
and we show that with this parametrization discretization errors -- or
computation time -- can be
reduced dramatically compared to the standard discretization.
}{}{}

\vspace*{10pt}


\setcounter{section}{2}
\setcounter{equation}{0}

\vspace*{1pt}\textlineskip      
\section{Introduction}          
\vspace*{-0.5pt}
\noindent
It is a standard procedure to introduce a discretized space-time in the
numerical study of physical problems. This is the case in quantum field
theories where the
numerical simulations are performed on a space-time lattice, or in classical
field theories (hydrodynamics, electrodynamics etc.), where the corresponding
partial differential equations can be discretized. The underlying mesh has a
finite lattice spacing $a$ introducing a discretization error whose size
depends on $a/\xi$, where $\xi$ is a typical length scale of the problem. This
systematical error is influenced by the discretized form of the differential
operators. For example, the standard nearest-neighbor discretization of the
Laplace operator has an ${\mathcal{O}} (a^2)$ error. By adding additional
couplings this error can be reduced to ${\mathcal{O}}(a^4)$, which makes it
easier to perform the continuum limit $a\rightarrow 0$. In principle, the
discretization error can be eliminated order by order this way, but the
resulting Laplace operator would not have much practical value in general: it
would 
become a broad, non-local operator.

The notion of locality will play an important role in the following. We shall
call a discretized differential operator local if the size of the region $R$
where the 
couplings are significantly different from zero goes to zero relative to the
typical length scale $\xi$ of the problem in the continuum limit:
$R/\xi\rightarrow 0$ $(a\rightarrow 0)$. The standard nearest-neighbor
discretization of the Laplace operator is obviously local.
Similarly, a
lattice difference operator whose couplings decay exponentially $\propto
e^{-\gamma r}$ with the
distance $r$ between the connected points, where
$\gamma a={\mathcal{O}}(1)$, is local also. In this case $R/\xi\propto
1/(\gamma \xi)\propto a/\xi\rightarrow 0$ in the continuum limit.

\textheight=7.7truein           
\setcounter{footnote}{0}
\renewcommand{\thefootnote}{\alph{footnote}}

Is it possible to construct ``perfect'' discretized differential operators,
i.e.~operators which are local and lead to differential equations on the
lattice whose solution is free of discretization errors even if $a/\xi$ is not
small and thus the resolution is bad? The answer is yes. The problem is
related to the existence of perfect lattice actions of classical field
theories \cite{hasenfratz_niedermayer:94}. The Euler-Lagrange equations corresponding to such actions
are perfect difference equations in the sense discussed above. Their existence
and the technical way of constructing them follows directly from Wilson's
renormalization group (RG) theory \cite{wilson_kogut:74}. Perfect classical actions were
constructed and tested not only for free field theories like the scalar field \cite{bietenholz:99}, but also for such
non-trivial cases like the non-linear $\sigma$-model \cite{blatter_burkhalter:95,blatter_burkhalter:96},
Yang-Mills gauge theory
\cite{degrand_hasenfratz:95,bietenholz_wiese:96} or the Schwinger
model \cite{bietenholz_wiese:96-2,lang_pany:98,farchioni_laliena:98}.
These results might open new paths in the numerical study of 
partial differential equations. The first
attempt in this direction has been made by Katz and 
Wiese for fluid dynamics \cite{katz_wiese:97}. Although the latter problem
seems to be
simpler than those treated before, there are new difficulties
here --- among them the influence of non-trivial boundary conditions. Consider the following simple
example: the potential energy of a massive membrane fixed to a frame at its
boundary. This is a good problem for
testing, as it can be solved analytically for a square frame of size $L\times
L$ 
and therefore the exact solution is known. The standard procedure to
solve this problem numerically is to define a lattice and calculate the
amplitude of the membrane at the lattice points using the standard
discretization of the Laplace operator. However, if $a/L$ is not very small,
the results will be far from  
the exact value due to the large discretization error. To get
acceptable results, one needs to go to very fine 
lattices, that is to very high resolution. On the other hand, as we mentioned, the perfect
lattice Laplace operator gives the exact value for the quantity we are looking
for at any 
lattice size, at any resolution, even when the lattice consists of only one
point.
For the boundary conditions mentioned above, the perfect Laplace operator can
even be found analytically.

So far, this is theoretical. For practical calculations, one should use a
truncated operator which is easy to handle. For example, one might consider
only nearest-neighbor and next-to-nearest-neighbor couplings between 
lattice points. A boundary will then influence these couplings. Such a
truncation only works when the neglected couplings are small, that is
when the operator is local. If the influence of the
boundary is also local, a
parametrization could be constructed which can be used for arbitrary boundary
shapes. In this paper, we construct and
test such a parametrization. 
Let us summarize the problems addressed:
\begin{itemize}
  \item How do non-trivial boundary conditions affect the perfect Laplace operator?
  \item Can we truncate and parametrize a perfect difference operator in a
    way that it is useful for practical purposes?
  \item Is the improvement we achieve for specific results when using a
    parametrized perfect operator worth the effort? 
\end{itemize}
As this is a field practically untouched, we turn to the basic problems
first. We will see that there the concept of perfect actions leads to
excellent results.

\vspace*{1pt}\textlineskip   
\section{Renormalization Group Transformations and Perfect Actions}
\vspace*{-0.5pt}
\noindent
Let us first give a brief review of the construction of perfect actions.
Consider a theory described by an action ${\mathcal{A}}(\phi_n)$,
where $\phi_n$ is the $d$-dimensional field variable at the lattice point
$n=(n_1,\dots,n_d)$. On this lattice, form blocks of $2^d$ points each. For every
block, define a blocked field variable
\begin{equation}
\chi_{n_B} = b\frac{1}{2^d}\sum_{n\in n_B} \phi_n.
\end{equation}
If the original lattice spacing has been $a$, the blocked field $\chi_{n_B}$
lives on a lattice with spacing $2a$. A Renormalization Group transformation
(RGT) step leads to a new action 
${\mathcal{A}}^\prime(\chi)$ for the blocked variable by integration over the
original field variables $\phi_n$:
\begin{equation} \label{basic_rgt}
e^{-{\mathcal{A}}^\prime(\chi)} = \prod_n\int d\phi_n \prod_{n_B}
\delta(\chi_{n_B} - b\frac{1}{2^d}\sum_{n\in n_B} \phi_n)
e^{-{\mathcal{A}}(\phi)} .
\end{equation}
It remains to rescale the unit of length in order to keep physical length
scales unchanged. After such an RGT step, our theory still
describes the same 
long-distance behaviour, i.e.~the same physics. But we have reduced the
number of degrees of freedom, that is the number of space-time variables, by a
factor of 
$2^d$. On the other hand, the form of the action has changed. Iterating
this transformation, one gets coarser and coarser lattices without adding
new discretization errors.

An interesting property of RG transformations is the occurrence of fixed
points. 
Most generally, the action of a theory consists of all kinds of interactions
and can be written as a sum of interaction
terms $\theta_i(\phi),\ i=1,2,\dots$
\begin{equation}
{\mathcal{A}}(\phi) = \sum_i K_i\theta_i(\phi),
\end{equation}
where $K_i$ are the respective coupling constants. As we said, a RG
transformation changes the form of the action, so repeated
transformations generate a flow in coupling constant space
\begin{equation}
\{K_i^{(1)}\} \rightarrow \{K_i^{(2)}\} \rightarrow
\{K_i^{(3)}\} \rightarrow \dots\enspace.
\end{equation}
A fixed point occurs if the set of coupling constants remains
unchanged under a RG transformation $\{K_i^{(n)}\} = \{K_i^{(n+1)}\}\doteq
\{K_i^*\}$. 
The fixed point action, which we designate by an asterisk, is then defined as 
\begin{equation}
{\mathcal{A}}^*(\phi) = \sum_{i} K_i^*\theta_i(\phi),
\end{equation}
and depends on the explicit form of the RG
transformation. The fixed point action has the beautiful property of being
classically 
perfect: It reproduces all the important classical properties of
the continuum action \cite{hasenfratz_niedermayer:94}.

Now let us consider an example: The perfect
Laplace operator $\Delta^*$ on a $d$-dimensional space-time without boundaries can be 
calculated from the fixed point action of a free real scalar field with
the continuum action
\begin{equation} \label{contfreeaction}
{\mathcal{A}}(\phi) = \frac{1}{2}\int\! d^dx\ 
\partial_\mu\phi(x)\partial_\mu\phi(x).
\end{equation}
The equation of motion for this action is the Laplace equation. A general
discretization of the action (\ref{contfreeaction}) contains terms which couple
the field at one lattice site to the field at another one:
\begin{equation} \label{quadratic_form}
{\mathcal{A}}(\phi) = \frac{1}{2} \sum_{n,r}\phi_n \rho(r) \phi_{n+r},
\end{equation}
with the coupling constants $\rho(r)$, $r=(r_1,\dots,r_d)$. For the standard Laplacian in two
dimensions, $\rho(0)=4$ and $\rho(r)=-1\ (\forall |r|=1)$. 
We generalize the block transformation in (\ref{basic_rgt}) and write
\begin{equation} \label{generalrgt}
  c \cdot e^{-{\mathcal{A}}^\prime(\chi)} = \prod_n\int\! d\phi_n \
  e^{-{\mathcal{A}}(\phi) - {\mathcal{T}}(\chi,\phi)} .
\end{equation}
where $c$ is a normalization constant and ${\mathcal{T}}(\chi,\phi)$ is the blocking kernel
\begin{equation} \label{gauss_block_kernel}
{\mathcal{T}}(\chi,\phi) = 2\kappa \sum_{n_B}(\chi_{n_B} -
  b\cdot\frac{1}{2^d}\sum_{n\in n_B} \phi_n)^2.
\end{equation}
The parameter $\kappa$ gives us the
possibility to optimize the fixed point action for locality. For
$\kappa \rightarrow\infty$, this RG transformation goes over to the one in
Eq.~(\ref{basic_rgt}). 
The fixed point of Eq.~(\ref{generalrgt}) can be calculated analytically
\cite{bell_wilson:74,bell_wilson:75}. In momentum space, the inverse of $\rho^*(q)$ is
\begin{equation} \label{inf_fp_prop}
\frac{1}{\tilde\rho^*(q)} = \sum_{l\in {\mathbf Z}^d} \frac{1}{(q+2\pi l)^2}
\prod_\mu \frac{\sin^2(\frac{q_\mu}{2}+\pi l_\mu)}{(\frac{q_\mu}{2}+\pi
  l_\mu)^2} + \frac{1}{3\kappa}.
\end{equation}
The fixed point Laplacian $(\Delta^*\phi)_n = -\sum_r\rho^*(r)\phi_{n+r}$ is
then calculated 
numerically by Fourier transforming $\tilde\rho^*(q)$.
A thorough examination of different blocking kernels \cite{ruefenacht:98} has shown that the kernel
(\ref{gauss_block_kernel}) with $\kappa=2$ gives very good results in terms of
locality: The couplings $\rho(r)$ decay exponentially $\propto e^{-\gamma r}$ with
a large decay coefficient $\gamma\approx 3.5$.
For $d=2$, we get the translationally invariant perfect Laplacian on a plane,
which we will use in the next section.

\section{Perfect Laplacian for a Square Boundary}
\noindent
The perfect Laplace operator calculated above only holds in the absence of
boundaries. So let us now introduce non-trivial boundary conditions: Restrict the field $\phi$
to a square area in two dimensions with $\phi=0$ 
on the boundaries. In the presence of boundaries,
the perfect Laplacian $\rho^*(n,n^\prime)$ will no longer be translation
invariant. In this section we present two ways of  
calculating the perfect Laplacian for these boundary
conditions. First we explicitly perform the RG transformation, 
leading to an analytical expression for the perfect Laplacian, and second we
show how to find  
$\rho^*(n,n^\prime)$ as a function of the translationally invariant perfect
Laplacian $\rho^*(r)$, which is already known from (\ref{inf_fp_prop}). The
latter procedure will lead to an 
elegant way of avoiding possible normalization problems when truncating the
couplings.

\vspace*{1pt}\textlineskip   
\subsection{Explicit RG Transformation}
\begin{figure}
\begin{center}
\setlength{\unitlength}{0.5cm}
\begin{picture}(8,9)
\put(0,0){\framebox(8,8)}
\multiput(1,7.9)(1,0){7}{\line(0,1){0.2}}
\multiput(-0.1,1)(0,1){7}{\line(1,0){0.2}}
\put(-1.3,-0.5){\makebox(1,1)[r]{$0$}}
\put(-1.3,0.5){\makebox(1,1)[r]{$a$}}
\put(-1.3,1.5){\makebox(1,1)[r]{$2a$}}
\put(-1.3,2.5){\makebox(1,1.5)[r]{\vdots}}
\put(-1.3,7.5){\makebox(1,1)[r]{$Na$}}
\put(4,8.5){\vector(1,0){4}}
\put(4,8.5){\vector(-1,0){4}}
\put(2,8.8){\makebox(4,1)[b]{$L=Na$}}
\multiput(0.5,0.5)(0,1){8}
  {\multiput(0,0)(1,0){8}{\circle*{0.12}} }
\put(1.5,3){\vector(0,1){0.5}}
\put(1.5,3){\vector(0,-1){0.5}}
\put(1.7,2.8){$a$}
\linethickness{0.04mm}
\multiput(0.3,0.3)(0,2){4}
  {\multiput(0,0)(2,0){4}{\dashbox{0.2}(1.4,1.4)} }
\end{picture} 
\end{center}
\caption{The $N\times N$-lattice in an area with square boundaries. The
  blocking kernel (\ref{gauss_block_kernel}) connects
  the four fine lattice points inside a dashed box to one coarse lattice point
  in the center of the box.}
  \label{latticefig} 
\end{figure}
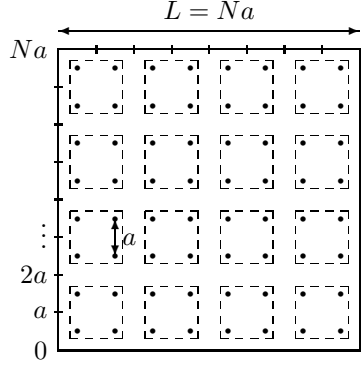

\vspace*{-0.5pt}
\noindent
Consider a square box with side length $L$ in $d=2$. In the continuum, the
action of a free scalar field is
\begin{equation} \label{freeact}
  {\mathcal{A}}(\phi)  =  
  \frac{1}{2}\int_{0}^{L}d^2x \enspace \partial_{\mu}\phi(x)\,
  \partial_{\mu}\phi(x).
\end{equation}
For the calculation of the fixed point action, we will work in Fourier
space. Our ansatz is to use a Fourier transformation 
\begin{equation}
  \phi(x)  =  \frac{1}{L^2} \sum_{q} \Psi_q(x)\,\tilde\phi(q),
\end{equation}
with orthonormal basis functions $\Psi_{q}(x) = 2\sin(q_1x_1)\sin(q_2x_2)$
which ensure that
the field $\phi$ fulfills the boundary conditions. 
The momentum variable $q=(q_1,q_2)$ takes the discrete set of values 
$q_i=k_i\pi/L $ with $  k_i=1,2,\dots,\infty$ ($i=1,2$).
In momentum space, the action (\ref{freeact}) reads
\begin{equation} \label{momact}
{\mathcal{A}}(\tilde\phi) = \frac{1}{2L^2} \sum_q q^2 \tilde\phi(q)
\tilde\phi(q).
\end{equation}
As shown in Fig.~\ref{latticefig}, we define the lattice points $n_i$
at half-integer values of
the continuum space variable $x_i = (n_i+1/2)a$ with $0\!\le\! x_i\!<\!L=Na$.
Performing the RG transformation (\ref{generalrgt}), (\ref{gauss_block_kernel}) on this lattice, we arrive at a
coarser lattice with lattice unit $2a$, defined in the middle of the dashed
blocks in Fig.~\ref{latticefig}.
Consider the correlation function
\begin{equation}
  \langle \phi_{n^\prime}\phi_{n^{\prime\prime}} \rangle  
  = \frac {1}{Z} \prod_n\int d\phi_n e^{-{\mathcal{A}}(\phi)}
  \phi_{n^\prime}\phi_{n^{\prime\prime}}, 
\end{equation}
where $Z$ is the partition function and assume
${\mathcal{A}}={\mathcal{A}}^*$. From Eq.~(\ref{generalrgt}) follows that the correlation
functions on the coarse and fine lattices are related by
\begin{equation}
  \langle \chi_{n_B}\chi_{n_B^{\prime}} \rangle  =
  \biggl(\frac{1}{4}\biggr)^2 \sum_{n\in n_B} \sum_{n^{\prime}\in n_B^{\prime}}
  \langle \phi_n\phi_{n^{\prime}} \rangle  + \frac{1}{4\kappa} \delta_{n_B
  n_B^{\prime}}.
\end{equation}
Iterating this transformation an infinite number of times
leads to a relation between the correlation functions on the original lattice
with 
lattice unit $a$ and the one in the continuum. (As we
need no longer two lattice field variables, we replace the coarse
lattice 
variable $\chi_{n_B}$ by our standard lattice notation $\phi_n$ again.) 
\begin{equation} \label{contcorr}
  \langle \phi_n\phi_{n^\prime} \rangle = \int_0^1 \!\!d^2x \int_0^1 \!\!d^2
    x^\prime 
    \langle \phi((n+x)a) \phi((n^{\prime}+x^{\prime})a) \rangle
  + \frac{1}{3\kappa}\delta_{nn^\prime}.
\end{equation}
On the right hand side, we insert the continuum free field propagator
\begin{equation} \label{freeprop}
  \langle \phi(x)\phi(y) \rangle  =  \frac{1}{L^2}\sum_q
  \Psi_q(x)\Psi_q(y)\frac{1}{q^2},
\end{equation}
which follows from (\ref{momact}), with $x_i,y_i\in(0,L)$. Setting the lattice unit $a=1$
and performing the definite integration over $x$ and $x^\prime$ gives
\begin{equation} \label{corr2}
  \langle \phi_{n}\phi_{n^{\prime}} \rangle  =
  \frac{4}{N^2} \sum_q \prod_{i=1}^{2} \sin (q_i(n_i+\frac{1}{2})) \sin
  (q_i(n_i^{\prime}+\frac{1}{2})) 
  \prod_{i=1}^{2} \frac{(\sin \frac{q_i}{2})^2} {(\frac{q_i}{2})^2}
  \frac{1}{q^2} 
  + \frac{1}{3\kappa}\delta_{nn^{\prime}}.
\end{equation}
After a few algebraic steps, we can bring this into the following form:
\begin{equation} \label{fp_prop_res}
  \langle \phi_{n}\phi_{n^{\prime}} \rangle  =
  \frac{1}{N^2} \sum_Q \Psi_Q(n) \Psi_Q(n^{\prime}) \cdot
  \sum_{l=-\infty}^{\infty} \frac{1}{(Q+2\pi l)^2}
  \prod_{i=1}^{2}\frac{\sin^2\frac{Q_i}{2}}{(\frac{Q_i}{2}+\pi l_i)^2}
  + \frac{1}{3\kappa} \delta_{nn^{\prime}},
\end{equation}
with $\Psi_Q(n) = \xi_{Q_1}(n_1)\cdot \xi_{Q_2}(n_2)$, and
\begin{equation} \label{fq}
  \xi_{Q_i}(n_i) = \left\{
    \begin{array}{ll}
    \sqrt{2}\sin (Q_i(n_i+\frac{1}{2})) &,Q_i\ne\pi, \\
    \sin (Q_i(n_i+\frac{1}{2})) &,Q_i=\pi.
    \end{array} \right.
\end{equation}
The new momentum variable $Q$ is restricted to the Brillouin zone and takes
the values $Q_i=k_i\pi/N$, $k_i=1,\dots,N$.
The result (\ref{fp_prop_res}) provides us the fixed point propagator in
momentum space 
\begin{equation} \label{fppropmom}
  \frac{1}{\rho^{*}(Q)} = \sum_{l=-\infty}^{\infty} \frac{1}{(Q+2\pi l)^2}
  \prod_{i=1}^{2}\frac{\sin^2\frac{Q_i}{2}}{(\frac{Q_i}{2}+\pi l_i)^2}
  + \frac{1}{3\kappa}.
\end{equation}
Perform the Fourier transformation
\begin{equation} \label{perf_lapl_conf_space}
  \rho^{*}(n,n^{\prime}) =
  \frac{1}{N^2} \sum_Q \Psi_Q(n) \Psi_Q(n^{\prime}) \cdot \rho^{*}(Q),
\end{equation}
to get the couplings of the fixed point action
$  {\mathcal{A}}^*(\phi)=1/2\sum_{n,n^{\prime}}
    \rho^{*}(n,n^{\prime})\phi_n\phi_{n^{\prime}}$ in configuration space
($n_i=0,1,\dots,N-1$). 
In contrast to the situation on an unbounded lattice, $\rho^*(n,n^\prime)$ is
no 
longer translationally invariant; it does
not depend only on the distance $r=n^\prime-n$ as the couplings in
Eq.~(\ref{quadratic_form}). Consequently, we have a different $\rho^*(n,r)$ for
every lattice 
point $n$. On the other hand we
expect --- and it is really the case, as we will show --- that away from the
boundaries the difference between 
$\rho^*(n,r)$ and the perfect Laplacian of the translationally invariant
case $\rho^*(r)$ goes to zero exponentially with the distance from the
boundary, which 
will be a crucial property to construct a parametrization.

\begin{table}

\footnotesize
\begin{center}

\setlength{\unitlength}{0.1cm}
\begin{picture}(10,10)
  \put(0,0){\dashbox(10,10)}
  \put(5,5){\circle{1.5}}
  \multiput(1,1)(0,2){5}{\multiput(0,0)(2,0){5}{\circle*{0.4}}}
  \put(4,-2.5){$\rho_1$}
\end{picture}
\hspace{0.5cm}
\begin{tabular}{|r||r|r|r|r|r|} \hline
 $\rho_1(r_1,r_2)$  & $r_1=-2$ & $r_1=-1$ & $r_1=0$ & $r_1=1$ & $r_1=2$  \\
 \hline \hline 
 $r_2=2$ &  0.00162 & -0.00068 & -0.00200 & -0.00068 &  0.00162 \\ \hline
 $r_2=1$ & -0.00068 & -0.19024 & -0.61773 & -0.19024 & -0.00068 \\ \hline
 $r_2=0$ & -0.00200 & -0.61773 &  3.23881 & -0.61773 & -0.00200 \\ \hline
 $r_2=-1$ & -0.00068 & -0.19024 & -0.61773 & -0.19024 & -0.00068 \\ \hline
 $r_2=-2$ &  0.00162 & -0.00068 & -0.00200 & -0.00068 &  0.00162 \\ \hline
\end{tabular}

\vspace{3mm}

\begin{picture}(10,10)
  \put(0,0){\dashbox(10,10)}
  \put(0,8){\line(1,0){10}}
  \put(5,5){\circle{1.5}}
  \multiput(1,1)(0,2){4}{\multiput(0,0)(2,0){5}{\circle*{0.4}}}
  \put(4,-2.5){$\rho_{2}$}
\end{picture}
\hspace{0.5cm}
\begin{tabular}{|r||r|r|r|r|r|} \hline
 $\rho_{2}(r_1,r_2)$  & $r_1=-2$ & $r_1=-1$ & $r_1=0$ & $r_1=1$ & $r_1=2$  \\
 \hline \hline 
 $r_2=2$ & 0.00000 &  0.00000 &  0.00000 &  0.00000 &  0.00000\\ \hline
 $r_2=1$ &-0.00230 & -0.18956 & -0.61573 & -0.18956 & -0.00230\\ \hline
 $r_2=0$ &-0.00200 & -0.61773 &  3.23881 & -0.61773 & -0.00200\\ \hline
 $r_2=-1$& -0.00068&  -0.19024&  -0.61773&  -0.19024&  -0.00068\\ \hline
 $r_2=-2$&  0.00162&  -0.00068&  -0.00200&  -0.00068&   0.00162\\ \hline
\end{tabular}

\vspace{3mm}

\begin{picture}(10,10)
  \put(0,0){\dashbox(10,10)}
  \put(0,6){\line(1,0){10}}
  \put(5,5){\circle{1.5}}
  \multiput(1,1)(0,2){3}{\multiput(0,0)(2,0){5}{\circle*{0.4}}}
  \put(4,-2.5){$\rho_{3}$}
\end{picture}
\hspace{0.5cm}
\begin{tabular}{|r||r|r|r|r|r|} \hline
 $\rho_{3}(r_1,r_2)$  & $r_1=-2$ & $r_1=-1$ & $r_1=0$ & $r_1=1$ & $r_1=2$  \\
 \hline \hline 
 $r_2=2$ &  0.00000 &  0.00000 &  0.00000 &  0.00000 &  0.00000\\ \hline
 $r_2=1$ &  0.00000 &  0.00000 &  0.00000 &  0.00000 &  0.00000\\ \hline
 $r_2=0$ & -0.00132 & -0.42749 &  3.85654 & -0.42749 & -0.00132\\ \hline
 $r_2=-1$ & -0.00230 & -0.18956 & -0.61573 & -0.18956 & -0.00230\\ \hline
 $r_2=-2$ &  0.00162 & -0.00068 & -0.00200 & -0.00068 &  0.00162\\ \hline
\end{tabular}

\vspace{3mm}

\begin{picture}(10,10)
  \put(0,0){\dashbox(10,10)}
  \put(0,8){\line(1,0){8}}
  \put(8,0){\line(0,1){8}}
  \put(5,5){\circle{1.5}}
  \multiput(1,1)(0,2){4}{\multiput(0,0)(2,0){4}{\circle*{0.4}}}
  \put(4,-2.5){$\rho_{4}$}
\end{picture}
\hspace{0.5cm}
\begin{tabular}{|r||r|r|r|r|r|} \hline
 $\rho_{4}(r_1,r_2)$  & $r_1=-2$ & $r_1=-1$ & $r_1=0$ & $r_1=1$ & $r_1=2$  \\
 \hline \hline 
 $r_2=2$ & 0.00000 &  0.00000 &  0.00000 &  0.00000 & \ 0.00000\\ \hline
 $r_2=1$ &-0.00230 & -0.18956 & -0.61573 & -0.18726 &  0.00000\\ \hline
 $r_2=0$ &-0.00200 & -0.61773 &  3.23881 & -0.61573 &  0.00000\\ \hline
 $r_2=-1$& -0.00068&  -0.19024&  -0.61773&  -0.18956&   0.00000\\ \hline
 $r_2=-2$&  0.00162&  -0.00068&  -0.00200&  -0.00230&   0.00000\\ \hline
\end{tabular}

\vspace{3mm}

\begin{picture}(10,10)
  \put(0,0){\dashbox(10,10)}
  \put(0,6){\line(1,0){8}}
  \put(8,0){\line(0,1){6}}
  \put(5,5){\circle{1.5}}
  \multiput(1,1)(0,2){3}{\multiput(0,0)(2,0){4}{\circle*{0.4}}}
  \put(4,-2.5){$\rho_{5}$}
\end{picture}
\hspace{0.5cm}
\begin{tabular}{|r||r|r|r|r|r|} \hline
 $\rho_{5}(r_1,r_2)$  & $r_1=-2$ & $r_1=-1$ & $r_1=0$ & $r_1=1$ & $r_1=2$  \\
 \hline \hline 
 $r_2=2$ & 0.00000 &  0.00000 &  0.00000 &  0.00000 & \ 0.00000 \\ \hline
 $r_2=1$ & 0.00000 &  0.00000 &  0.00000 &  0.00000 &  0.00000 \\ \hline
 $r_2=0$ &-0.00132 & -0.42749 &  3.85654 & -0.42617 &  0.00000 \\ \hline
 $r_2=-1$ &-0.00230 & -0.18956 & -0.61573 & -0.18726 &  0.00000 \\ \hline
 $r_2=-2$ & 0.00162 & -0.00068 & -0.00200 & -0.00230 &  0.00000 \\ \hline
\end{tabular}

\vspace{3mm}

\begin{picture}(10,10)
  \put(0,0){\dashbox(10,10)}
  \put(0,6){\line(1,0){6}}
  \put(6,0){\line(0,1){6}}
  \put(5,5){\circle{1.5}}
  \multiput(1,1)(0,2){3}{\multiput(0,0)(2,0){3}{\circle*{0.4}}}
  \put(4,-2.5){$\rho_{6}$}
\end{picture}
\hspace{0.5cm}
\begin{tabular}{|r||r|r|r|r|r|} \hline
 $\rho_{6}(r_1,r_2)$  & $r_1=-2$ & $r_1=-1$ & $r_1=0$ & $r_1=1$ & $r_1=2$  \\
 \hline \hline 
 $r_2=2$ &  0.00000 &  0.00000 &  0.00000 & \ 0.00000 & \ 0.00000\\ \hline
 $r_2=1$ &  0.00000 &  0.00000 &  0.00000 &  0.00000 &  0.00000\\ \hline
 $r_2=0$ & -0.00132 & -0.42617 &  4.28403 &  0.00000 &  0.00000\\ \hline
 $r_2=-1$ & -0.00230 & -0.18726 & -0.42617 &  0.00000 &  0.00000\\ \hline
 $r_2=-2$ &  0.00162 & -0.00230 & -0.00132 &  0.00000 &  0.00000\\ \hline
\end{tabular}
\end{center}
\normalsize

\caption{The normalized, truncated couplings of the perfect Laplacian for
  points far from boundaries
 ($\rho_1$), for points near a wall ($\rho_2,\rho_3$) and near a
  convex corner ($\rho_4,\rho_5,\rho_6$). In the small image on the left, the
  lattice point $n$ is denoted by a circle, and all the sites in the
  2-hypercube around $n$ which lie inside the boundary are shown as small
  dots. The table on the right shows the values of the $(r_1,r_2)$-coupling
  for the lattice point $n$. The couplings for $\rho_2$--$\rho_6$ are made up
 from $\rho_1$ with the construction explained in the text.}\label{table_cpl1}
\end{table}

\vspace*{1pt}\textlineskip   
\subsection{Construction from Symmetry Properties}
\vspace*{-0.5pt}
\noindent
There is another, more elegant way to find the perfect Laplacian for a square
region with zero boundary conditions in $d=2$. The
translationally invariant perfect Laplacian $\rho^*(r)$ (\ref{inf_fp_prop}) is
already known, and
we show that this knowledge can be used to solve the problem with these non-trivial
boundary conditions.
As a boundary, consider a single wall at $x_1=0$ first where
the field $\phi$ has to vanish. We extend the field beyond the boundary and
introduce the condition 
\begin{equation}
  \phi(-x_1,x_2)=-\phi(x_1,x_2),
\end{equation}
which implies $\phi=0$ on the boundary. For the lattice field, 
we have
$\phi_{-n_1-1, n_2} = -\phi_{n_1,n_2}$,
as the lattice points $n$ are at half-integer values of
$x=n+1/2$. Using this symmetry relation and the symmetry property
$\rho^*(-r_1,r_2)=\rho^*(r_1,r_2)$, we rewrite the perfect Laplace equation on
the unbounded lattice into an equation on the right halfplane:
\begin{eqnarray} \label{eq_split_wall}
  \sum_r\rho^*(r)\phi_{n+r} &=& \sum_{r_2} \Bigl[ \sum_{r_1\geq 0}
  \rho^*(r-n)\phi_{r} + \sum_{r_1<0}
  \rho^*(r-n)\phi_{r} \Bigr] \\
 &=& \sum_{r_2}\sum_{r_1\geq 0} \bigl[ \rho^*(r-n) - \rho^*(r_1+n_1+1,r_2-n_2)
  \bigr] \phi_{r}. \nonumber
\end{eqnarray}
This defines a perfect Laplace operator with $\phi=0$ on the second axis,
that is the perfect Laplacian near a wall
\begin{equation} \label{eq_lapwall}
  \rho^*(n,r) = \rho^*(r) - \rho^*(r_1+2n_1+1,r_2),
\end{equation}
where $n_1$ is the distance from the wall.
The variables $n$ and $n^\prime=n+r$ are here restricted to the area inside
the boundary which is the positive halfplane in the first coordinate.

Now consider a boundary of two walls at $x_1=0$ and $x_2=0$ forming a corner. 
In the Laplace equation, the sum over the lattice points can be split into
sums over the four quadrants of the plane. Proceeding as in the above case for
a wall, we get the perfect Laplacian 
\begin{eqnarray} \label{eq_lapcorner}
  \rho^*(n,r) = \rho^*(r) - \rho^*(r_1+2n_1+1,r_2) - \rho^*(r_1,r_2+2n_2+1) \\
  + \rho^*(r_1+2n_1+1,r_2+2n_2+1), \nonumber
\end{eqnarray}
with both $n$ and $n+r$ lying in the first quadrant.

Finally, we may form a square boundary out of four walls at $x_1=0$,
$x_1=N$, $x_2=0$ and $x_2=N$. The field outside
the boundary is formally defined by periodically mirroring it at the boundary
with alternating sign
\begin{equation}
  \phi(x_1,x_2) = (-1)^{k_1+k_2}
  \phi\bigl((-1)^{k_1}x_1+2l_1N,(-1)^{k_2}x_2+2l_2N\bigr),
\end{equation}
for any $k_i=0,1$ and $l_i\in{\mathbf Z}$, ($i=1,2$).
The sum over the
whole plane then splits up into sums over $N$-squares
\begin{equation}
  \sum_r\rho^*(r)\phi_{n+r} = \sum_{i=1}^2\sum_{l_i\in{\mathbf Z}} \sum_{k_i=0}^1 \sum_{s_i=0}^{N-1} \rho(r-n)\phi_r,
\end{equation}
where the variable $r$ running over the lattice points is given by
$r_i = (-1)^{k_i} s_i-k_i+2l_iN$
for $i=1,2$. The perfect Laplacian consists of the infinite sum of
couplings 
\begin{equation}\label{eq_sym_sq}
\rho^*(n,r) = \sum_{i=1}^2\sum_{l_i\in{\mathbf Z}} \sum_{k_i=0}^1
(-1)^{k_1+k_2} 
\rho^*\bigl(r_1+2k_1n_1+k_1+2l_1N,r_2+2k_2n_2+k_2+2l_2N\bigr), 
\end{equation}
with both $n$ and $n+r$ lying inside the boundary. A check 
for a lattice with $N=1$ shows that the relation (\ref{eq_sym_sq}) with the
translationally invariant couplings (\ref{inf_fp_prop}) 
gives the same result $\rho^*(0,0)=4.95513$ as the explicit
RG transformation (\ref{fppropmom}), (\ref{perf_lapl_conf_space}).

\vspace*{1pt}\textlineskip   
\section{Perfect Laplacian near a Concave Corner}
\vspace*{-0.5pt}
\noindent
For boundary shapes where
the fixed point action is hard or impossible to find analytically, RG
transformations may be performed numerically. We performed such a numerical
RGT in order to find the couplings of 
the fixed point Laplacian near the concave corner of an L-shaped boundary,
that is a square with a smaller square cut out, in
$d=2$ (see Fig.~\ref{fig_cclat}). For quadratic actions and blocking kernels,
the RG transformation Eq.~(\ref{generalrgt})
can be written as a minimizing condition for the fine field $\phi$:
\begin{equation}\label{eq_minstep}
{\mathcal{A}}^\prime(\chi) = \min_\phi \left[{\mathcal{A}}(\phi) +
  {\mathcal{T}}(\chi,\phi) \right] + \mbox{const.}
\end{equation}
To get close to a fixed point, we have to iterate this RGT step. The
results of the previous step --- which are the couplings of the resulting
coarse
action --- are then used as an input for the next step, that is as a new
starting guess for the fine action ${\mathcal{A}}(\phi)$.
After ${\mathcal{O}}(20)$ iterations, we find a very close
approximation to the couplings of the fixed point action. 
\setlength{\unitlength}{2mm}
\begin{figure}
\begin{center}
\begin{picture}(32,32)
\put(0,0){\line(0,1){32}}
\put(0,0){\line(1,0){32}}
\put(32,0){\line(0,1){16}}
\put(0,32){\line(1,0){16}}
\put(16,16){\line(0,1){16}}
\put(16,16){\line(1,0){16}}
\put(8,8){\dashbox(16,16)}
\put(12,12){\dashbox(8,8)}
\put(10.275,8.275){\framebox(1.45,1.45)}
\put(11,9){\line(1,2){2}}
\multiput(12.275,12.275)(2,0){4}{\multiput(0,0)(0,2){2}{\framebox(1.45,1.45)}}
\multiput(12.275,16.275)(2,0){2}{\multiput(0,0)(0,2){2}{\framebox(1.45,1.45)}} 
\multiput(0.5,0.5)(1,0){32}{\multiput(0,0)(0,1){16}{\circle*{0.15}}}
\multiput(0.5,16.5)(1,0){16}{\multiput(0,0)(0,1){16}{\circle*{0.15}}}
\put(15.5,-1.5){$n_1$}
\put(-2.5,15.5){$n_2$}
\put(-0.8,0.1){\scriptsize$1$}
\put(-1.3,31.1){\scriptsize$N$}
\put(0.2,-1.2){\scriptsize$1$}
\put(31,-1.2){\scriptsize$N$}
\put(18,24){\line(2,1){2}}
\put(20.8,24.9){\makebox(20,4)[lb]{\scriptsize Resulting coarse field
    couplings}} 
\put(18,20){\line(2,1){2}}
\put(20.8,20.8){\makebox(20,4)[lb]{\scriptsize Fine couplings needed for
    input}} 
\end{picture}
\caption{The $N=32$ - lattice used to calculate the couplings near a concave
  corner. The couplings between the
  fine lattice points (shown as dots) have to be fixed. The RG
  transformation then gives the 
  couplings between the coarse lattice points (denoted by squares) inside the
  large dashed box, 
  which are used as an input for the fine field couplings inside the small
  dashed box in the next iteration.} \label{fig_cclat}
\end{center}
\end{figure}
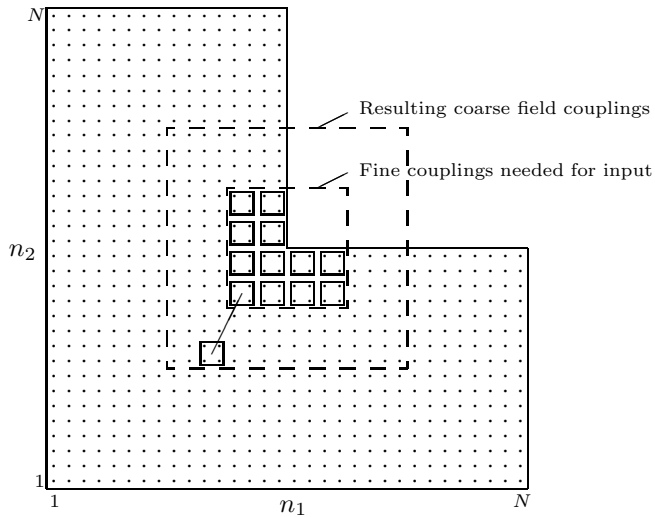

We worked on a lattice with $N=32$, as shown in
Fig.~\ref{fig_cclat}, and looked for the couplings of the coarse field inside
the large dashed box. Outside the small dashed box in Fig.~\ref{fig_cclat}, we
used in every
RGT step the parametrized couplings from Table \ref{table_cpl1} for the fine action. For the couplings inside the small dashed box, we used the standard
Laplacian in the first iteration, and afterwards the result of the previous
iteration. For every step, the coarse field couplings can be read out one by
one by choosing appropriate coarse field configurations as an input. Our
results for the fixed point couplings are listed in Table \ref{table_cpl2}. 
\begin{table}
\footnotesize
\begin{center}
\setlength{\unitlength}{0.1cm}
\begin{picture}(10,10)
  \put(0,0){\dashbox(10,10)}
  \put(8,8){\line(1,0){2}}
  \put(8,8){\line(0,1){2}}
  \put(5,5){\circle{1.5}}
  \multiput(1,1)(0,2){4}{\multiput(0,0)(2,0){5}{\circle*{0.4}}}
  \multiput(1,9)(0,2){1}{\multiput(0,0)(2,0){4}{\circle*{0.4}}}
  \put(4,-2.5){$\rho_{7}$}
\end{picture}
\hspace{0.5cm}
\begin{tabular}{|r||r|r|r|r|r|} \hline
 $\rho_{7}(r_1,r_2)$  & $r_1=-2$ & $r_1=-1$ & $r_1=0$ & $r_1=1$ & $r_1=2$  \\
 \hline \hline 
 $r_2=2$ &  0.00160 & -0.00072 & -0.00192 & -0.00206 &  0.00000 \\ \hline
 $r_2=1$ & -0.00071 & -0.19033 & -0.61787 & -0.19145 & -0.00206\\ \hline
 $r_2=0$ & -0.00190 & -0.61784 &  3.26307 & -0.61787 & -0.00192 \\ \hline
 $r_2=-1$ & -0.00071 & -0.19033 & -0.61784 & -0.19033 & -0.00072 \\ \hline
 $r_2=-2$ &  0.00160 & -0.00071 & -0.00190 & -0.00071 &  0.00160 \\ \hline
\end{tabular}

\vspace{3mm}

\begin{picture}(10,10)
  \put(0,0){\dashbox(10,10)}
  \put(6,8){\line(1,0){4}}
  \put(6,8){\line(0,1){2}}
  \put(5,5){\circle{1.5}}
  \multiput(1,1)(0,2){4}{\multiput(0,0)(2,0){5}{\circle*{0.4}}}
  \multiput(1,9)(0,2){1}{\multiput(0,0)(2,0){3}{\circle*{0.4}}}
  \put(4,-2.5){$\rho_{8}$}
\end{picture}
\hspace{0.5cm}
\begin{tabular}{|r||r|r|r|r|r|} \hline
 $\rho_{8}(r_1,r_2)$  & $r_1=-2$ & $r_1=-1$ & $r_1=0$ & $r_1=1$ & $r_1=2$  \\
 \hline \hline 
 $r_2=2$ &  0.00160 & -0.00071 & -0.00043 &  0.00000 &  0.00000 \\ \hline
 $r_2=1$ & -0.00071 & -0.19032 & -0.61625 & -0.18956 & -0.00233 \\ \hline
 $r_2=0$ & -0.00190 & -0.61787 &  3.26192 & -0.61787 & -0.00196 \\ \hline
 $r_2=-1$ & -0.00071 & -0.19033 & -0.61784 & -0.19033 & -0.00072 \\ \hline
 $r_2=-2$ &  0.00160 & -0.00071 & -0.00190 & -0.00071 &  0.00160 \\ \hline
\end{tabular}

\vspace{3mm}

\begin{picture}(10,10)
  \put(0,0){\dashbox(10,10)}
  \put(4,8){\line(1,0){6}}
  \put(4,8){\line(0,1){2}}
  \put(5,5){\circle{1.5}}
  \multiput(1,1)(0,2){4}{\multiput(0,0)(2,0){5}{\circle*{0.4}}}
  \multiput(1,9)(0,2){1}{\multiput(0,0)(2,0){2}{\circle*{0.4}}}
  \put(4,-2.5){$\rho_{9}$}
\end{picture}
\hspace{0.5cm}
\begin{tabular}{|r||r|r|r|r|r|} \hline
 $\rho_{9}(r_1,r_2)$  & $r_1=-2$ & $r_1=-1$ & $r_1=0$ & $r_1=1$ & $r_1=2$  \\
 \hline \hline 
 $r_2=2$ &  0.00160 &  0.00062 &  0.00000 &  0.00000 &  0.00000 \\ \hline
 $r_2=1$ & -0.00071 & -0.18948 & -0.61637 & -0.18963 & -0.00233 \\ \hline
 $r_2=0$ & -0.00192 & -0.61787 &  3.27774 & -0.61788 & -0.00196 \\ \hline
 $r_2=-1$ & -0.00072 & -0.19033 & -0.61784 & -0.19033 & -0.00072 \\ \hline
 $r_2=-2$ &  0.00160 & -0.00071 & -0.00190 & -0.00071 &  0.00160 \\ \hline
\end{tabular}

\vspace{3mm}

\begin{picture}(10,10)
  \put(0,0){\dashbox(10,10)}
  \put(2,8){\line(1,0){8}}
  \put(2,8){\line(0,1){2}}
  \put(5,5){\circle{1.5}}
  \multiput(1,1)(0,2){4}{\multiput(0,0)(2,0){5}{\circle*{0.4}}}
  \multiput(1,9)(0,2){1}{\multiput(0,0)(2,0){1}{\circle*{0.4}}}
  \put(4,-2.5){$\rho_{10}$}
\end{picture}
\hspace{0.5cm}
\begin{tabular}{|r||r|r|r|r|r|} \hline
 $\rho_{10}(r_1,r_2)$  & $r_1=-2$ & $r_1=-1$ & $r_1=0$ & $r_1=1$ & $r_1=2$  \\
 \hline \hline 
 $r_2=2$ &  0.00021 &  0.00000 &  0.00000 &  0.00000 &  0.00000 \\ \hline
 $r_2=1$ & -0.00206 & -0.18964 & -0.61594 & -0.18965 & -0.00233 \\ \hline
 $r_2=0$ & -0.00196 & -0.61788 &  3.24479 & -0.61788 & -0.00196 \\ \hline
 $r_2=-1$ & -0.00072 & -0.19033 & -0.61784 & -0.19033 & -0.00072 \\ \hline
 $r_2=-2$ &  0.00160 & -0.00071 & -0.00190 & -0.00071 &  0.00160 \\ \hline
\end{tabular}

\vspace{3mm}

\begin{picture}(10,10)
  \put(0,0){\dashbox(10,10)}
  \put(6,6){\line(1,0){4}}
  \put(6,6){\line(0,1){4}}
  \put(5,5){\circle{1.5}}
  \multiput(1,1)(0,2){3}{\multiput(0,0)(2,0){5}{\circle*{0.4}}}
  \multiput(1,7)(0,2){2}{\multiput(0,0)(2,0){3}{\circle*{0.4}}}
  \put(4,-2.5){$\rho_{11}$}
\end{picture}
\hspace{0.5cm}
\begin{tabular}{|r||r|r|r|r|r|} \hline
 $\rho_{11}(r_1,r_2)$  & $r_1=-2$ & $r_1=-1$ & $r_1=0$ & $r_1=1$ & $r_1=2$  \\
 \hline \hline 
 $r_2=2$ & 0.00155  &-0.00206  &-0.00051 &  0.00000 &  0.00000 \\ \hline
 $r_2=1$ & -0.00068 & -0.18948 & -0.46147 & 0.00000 &  0.00000 \\ \hline
 $r_2=0$ & -0.00186 & -0.61625 &  3.36152 & -0.46147 & -0.00051 \\ \hline
 $r_2=-1$ & -0.00076&  -0.19145 & -0.61625 & -0.18948 & -0.00206 \\ \hline
 $r_2=-2$ & 0.00161 & -0.00076  &-0.00186 & -0.00068 &  0.00155 \\ \hline
\end{tabular}

\vspace{3mm}

\begin{picture}(10,10)
  \put(0,0){\dashbox(10,10)}
  \put(4,6){\line(1,0){6}}
  \put(4,6){\line(0,1){4}}
  \put(5,5){\circle{1.5}}
  \multiput(1,1)(0,2){3}{\multiput(0,0)(2,0){5}{\circle*{0.4}}}
  \multiput(1,7)(0,2){2}{\multiput(0,0)(2,0){2}{\circle*{0.4}}}
  \put(4,-2.5){$\rho_{12}$}
\end{picture}
\hspace{0.5cm}
\begin{tabular}{|r||r|r|r|r|r|} \hline
 $\rho_{12}(r_1,r_2)$  & $r_1=-2$ & $r_1=-1$ & $r_1=0$ & $r_1=1$ & $r_1=2$  \\
 \hline \hline 
 $r_2=2$ &  0.00021 &  0.00055 &  0.00000 &  0.00000 &  0.00000 \\ \hline
 $r_2=1$ &  0.00062 & -0.02666 &  0.00000 &  0.00000 &  0.00000 \\ \hline
 $r_2=0$ & -0.00043 & -0.46147 &  3.88365 & -0.42738 & -0.00121 \\ \hline
 $r_2=-1$ & -0.00206 & -0.18956 & -0.61637 & -0.18964 & -0.00233 \\ \hline
 $r_2=-2$ &  0.00156 & -0.00074 & -0.00179 & -0.00073 &  0.00155 \\ \hline
\end{tabular}

\vspace{3mm}

\begin{picture}(10,10)
  \put(0,0){\dashbox(10,10)}
  \put(2,6){\line(1,0){8}}
  \put(2,6){\line(0,1){4}}
  \put(5,5){\circle{1.5}}
  \multiput(1,1)(0,2){3}{\multiput(0,0)(2,0){5}{\circle*{0.4}}}
  \multiput(1,7)(0,2){2}{\multiput(0,0)(2,0){1}{\circle*{0.4}}}
  \put(4,-2.5){$\rho_{13}$}
\end{picture}
\hspace{0.5cm}
\begin{tabular}{|r||r|r|r|r|r|} \hline
 $\rho_{13}(r_1,r_2)$  & $r_1=-2$ & $r_1=-1$ & $r_1=0$ & $r_1=1$ & $r_1=2$  \\
 \hline \hline 
 $r_2=2$ & -0.00001 &  0.00000 &  0.00000 &  0.00000 &  0.00000 \\ \hline
 $r_2=1$ &  0.00055 &  0.00000 &  0.00000 &  0.00000 &  0.00000 \\ \hline
 $r_2=0$ & -0.00051 & -0.42738 &  3.85640 & -0.42757 & -0.00121 \\ \hline
 $r_2=-1$ & -0.00233 & -0.18963 & -0.61594 & -0.18965 & -0.00233 \\ \hline
 $r_2=-2$ &  0.00155 & -0.00073 & -0.00179 & -0.00073 &  0.00155 \\ \hline
\end{tabular}

\end{center}
\caption{The truncated couplings of the perfect Laplacian near a concave
 corner. These couplings are the results of the iterative blocking procedure
 and are therefore not normalized.}\label{table_cpl2}
\end{table}

\vspace*{1pt}\textlineskip   
\section{Parametrization for Boundaries of Arbitrary Shape}
\vspace*{-0.5pt}
\noindent
The main observation which allows us to construct a reasonable parametrization
for the perfect Laplacian that can be used to approximate arbitrarily shaped
boundaries is the fact that not only its couplings decay
exponentially with the distance, but also the effect of the boundary on the
couplings: 
According to Eq.~(\ref{eq_lapwall}), the difference between the
perfect Laplacian near a wall and the translationally invariant perfect
Laplacian on a plane without boundaries
\begin{equation}
\Delta\rho_n^*(r) \doteq \rho^*(n,r) - \rho^*(r) = - \rho^*(r_1+2n_1+1,r_2),
\end{equation}
decays with the distance from the wall $n_1$ at twice the decay rate of the FP
 Laplacian couplings. The value of the decay constant $\gamma$, which is
 defined by $\Delta\rho_n^*(r) \propto e^{-\gamma \cdot n_1}$, where $n_1$
 denotes the orthogonal distance from the wall --- or the diagonal distance
 from 
 the corner, respectively --- is listed for several couplings $r$ near a wall
 and near a corner in Table \ref{tab_gamma_bnd}.
\begin{figure}[htb]
\begin{center}
\psfrag{xlabel}[][]{\hspace{3mm}\vspace{-0.1cm}$n_1$}
\psfrag{ylabel}[][]{\hspace{-2.5cm}\vspace{-0.2cm}$\log\left|
    \frac{\rho^*(n,r)-\rho^*(r)}{\rho^*(r)}\right|$}
\psfrag{-30}{\hspace{-1.5mm}\scriptsize$-30$}
\psfrag{-25}{\hspace{-1.5mm}\scriptsize$-25$}
\psfrag{-20}{\hspace{-1.5mm}\scriptsize$-20$}
\psfrag{-15}{\hspace{-1.5mm}\scriptsize$-15$}
\psfrag{-10}{\hspace{-1.5mm}\scriptsize$-10$}
\psfrag{-5}{\hspace{-1.5mm}\scriptsize$-5$}
\psfrag{0}{\scriptsize$0$}
\psfrag{1}{\scriptsize$1$}
\psfrag{2}{\scriptsize$2$}
\psfrag{3}{\scriptsize$3$}
\psfrag{4}{\scriptsize$4$}
\psfrag{5}{\scriptsize$5$}
\psfrag{6}{\scriptsize$6$}
\psfrag{7}{\scriptsize$7$}
\psfrag{8}{\scriptsize$8$}
\psfrag{(0,1)-couplinggg}{\scriptsize $r=(0,1)$}
\psfrag{(1,0)-coupling}{\scriptsize $r=(1,0)$}
\psfrag{(1,1)-coupling}{\scriptsize $r=(1,1)$}
\includegraphics[width=8cm]{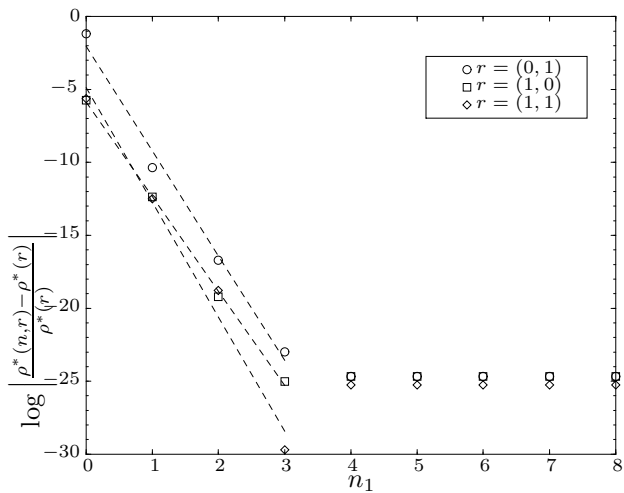}
\end{center}
\caption{The relative difference $\Delta\rho^*_n(r)$/$\rho^*(r)$ between three couplings
  near a 
  wall and the respective couplings of the perfect Laplacian on an infinite
  lattice. The couplings $\rho^*(n,r)$ 
  are calculated on a lattice with $N=17$ at the lattice sites
  $n=(n_1,(N+1)/2)$. $n_1$ 
  is therefore the distance from the wall in lattice units.  The difference
  decreases exponentially with the distance from the 
  wall. The slope $\gamma_w$ is listed in Table \ref{tab_gamma_bnd}. For 
  distances larger than 3 lattice units, the difference is already beyond the
  numerical accuracy.}
\label{plot_walldiff}
\end{figure}
\begin{table}[htb] 
\begin{center}
\begin{tabular}{|c|c|c|} \hline
$r$ & $\gamma^{(wall)}$ & $\gamma^{(corner)}$ \\ \hline 
(0,1) & 7.2 & 7.2 \\ 
(1,0) & 6.5 & 7.2 \\
(1,1) & 7.8 & 7.0 \\ 
(0,2) & 5.9 & 6.0 \\ \hline
\end{tabular}
\vspace{2mm}
\end{center}
\caption{The exponential decay rate $\gamma$ for the effect of the boundary
  on different couplings. The values are from linear least-squares fits to the
  first few data points in Figure \ref{plot_walldiff}.}
\label{tab_gamma_bnd}
\end{table}
This very strong exponential decay means that for
points a few lattice spacings away from walls, the presence of the boundary
influences the fixed point Laplacian at that point only negligibly, and
therefore we 
may parametrize the perfect Laplacian in a way that it can be used
for lattices with boundaries of an arbitrary shape.

In our parametrization, we take together points
with very similar $\rho^*(n,r)$ (e.g.\ all the points with a distance from the
boundary of more than one lattice unit) and use the same operator $\rho_m(r)$
for all these points. Thus, we classify the lattice points by their position
relative to the boundary (lower left corner, right wall, \dots) and define a
set of operators $\rho_m(r)$, $m=1,\dots,13$, where $m$ is the type of lattice point
as shown in Fig.~\ref{pointtypes}. Making this approximation, our set of
operators is no longer perfect, and so we have dropped the asterisk in the
notation.
\begin{figure}[htb]
\begin{center}
\setlength{\unitlength}{0.5cm}
\begin{picture}(4,4)
 \begin{small}
   \linethickness{0.2pt}
   \put(1,1){\framebox(1,1){1}} \put(2,1){\framebox(1,1){}}
   \put(3,1){\framebox(1,1){}} \put(1,2){\framebox(1,1){2}}
   \put(2,2){\framebox(1,1){4}} \put(3,2){\framebox(1,1){}}
   \put(1,3){\framebox(1,1){3}} \put(2,3){\framebox(1,1){5}}
   \put(3,3){\framebox(1,1){6}}
   \multiput(1,0)(1,0){4}{\line(0,1){4}}
   \multiput(0,1)(0,1){4}{\line(1,0){4}}

 \end{small}
\end{picture}
\end{center}
\caption{Point types used for the parametrization. The lattice points are
   classified by their relative position to the boundary. For every point type
   $m$ we define a different Laplace 
   operator $\rho_m(r)$ (see Table \ref{table_cpl1}). The definition of
   $\rho_7,\dots,\rho_{13}$ is shown in Table \ref{table_cpl2}.}\label{pointtypes}
\end{figure}

\vspace*{1pt}\textlineskip   
\subsection{Truncation and Normalization}
\vspace*{-0.5pt}
\noindent
For practical purposes, the perfect Laplacian must be truncated to a finite
number of couplings. The most primitive nearest-neighbor Laplace operator and
its simple improved 
versions have the basic property that they all approach the same exact
continuum result as the resolution is increased (universality). We have to
ensure that our approximate truncated perfect Laplace operator also has this
basic 
property. This can be achieved when imposing some conditions implied by
elementary principles on our parametrized action. Let us consider the case of an unbounded lattice first. Any discretized action
should describe the physical properties of the continuum action. For example,
the spectrum of the discretized action, which
is given by the poles of the propagator, has to go over to the spectrum of the
continuum action. In the continuum, the free field propagator
is $q^{-2}$, and the poles are at $q=(p,i|p|)$. The relativistic dispersion
relation in two euclidian dimensions is $E(k) = -iq_2 = |p|$. The
propagator $1/\rho(q)$ of a quadratic discretized action should have the
same poles, therefore the couplings have to fulfill the sum rule
\begin{eqnarray}
S_0 &=& \sum_{r_1,r_2} \rho(r_1,r_2) = 0.
\end{eqnarray}
Furthermore, we demand
that the lattice action takes the form of the continuum action for small
momenta $q$, that is 
$  \lim_{q\rightarrow 0} \rho(q) = q^2$.
This leads to the sum rule
\begin{equation}
  S_2 = \sum_{r_1,r_2} (r_1^2+r_2^2)\cdot\rho(r_1,r_2) = -4.
\end{equation}

When we have boundaries, the above considerations can't be used directly to
find sum 
rules. But with a different approach, we can at least find some conditions for
the couplings. Consider a wall at $x_1=0$. The field
$\phi(x)=x_1$ is a solution to the continuum Laplace equation and is zero at
the boundary. Therefore the lattice 
field $\Phi_n=n_1+1/2$ is a solution to the lattice Laplace equation,
and when inserting this solution we find a condition for the couplings $\rho_n(r)$:
\begin{equation} \label{eq_sumrule_w}
  S_1^{(w)} = \sum_{r_1,r_2}\rho_n(r_1,r_2) (r_1+\frac{1}{2})=0.
\end{equation}
For a corner, $\phi(x)=x_1x_2$ is a solution in the continuum. Hence, the
lattice field $\Phi_n=(n_1+1/2)(n_2+1/2)$ solves the lattice
Laplace equation, and the sum rule is
\begin{equation} \label{eq_sumrule_c}
  S_1^{(c)} = \sum_{r_1,r_2}\rho_n(r_1,r_2)
  (r_1+\frac{1}{2})(r_2+\frac{1}{2})=0. 
\end{equation}
While (\ref{eq_sumrule_w}) and (\ref{eq_sumrule_c}) are trivially fulfilled on
an unbounded lattice, 
they make sense as a normalizing condition for the couplings in the presence of a boundary.
We remark that Eq.~(\ref{eq_sumrule_c}) holds for both convex and concave
corners.

We construct our parametrization in the following manner: First we normalize
the truncated Laplacian for inner points $\rho_1$ --- which we take from
Eq.~(\ref{inf_fp_prop}) --- by setting the  
$(0,0)$-coupling properly to get $S_0=0$ and by rescaling all couplings to
ensure $S_2=-4$. Then we build the Laplace operators $\rho_2$,\dots,$\rho_6$
out of the already corrected couplings of $\rho_1$ with the 
help of the relations (\ref{eq_lapwall}) and (\ref{eq_lapcorner}). The
resulting couplings are listed in Table \ref{table_cpl1}. Together with the
couplings near a concave corner $\rho_7,\dots,\rho_{13}$, we have a
parametrization that can be used to approximate any two-dimensional shape.

\vspace*{1pt}\textlineskip 
\section{Testing the Parametrization}
\vspace*{-0.5pt}
\noindent
Let us first check the statement that the fixed point action is classically perfect
on an exactly solvable problem. Consider the free scalar field with a constant
source $f$:
\begin{equation} \label{class_act_src}
  {\mathcal{A}}(\phi)  =  
  \frac{1}{2}\int_{0}^{L}d^2x \enspace \partial_{\mu}\phi(x)\,
  \partial_{\mu}\phi(x) + f\int_0^Ld^2x \phi(x).
\end{equation}
We may interpret this as the action of a membrane which is fixed to $\phi=0$ at
the boundaries and has mass
density $f$. The total mass is then $F=fL^2$. 
The equation of motion to the action (\ref{class_act_src}) is the Poisson
equation 
$\Delta\phi(x)=f$. The Green's function to the Laplace operator which fulfills
the given boundary conditions is
\begin{equation} \label{green}
  G(x,y) = -\frac{1}{L^2} \sum_q \frac{\Psi_q(x) \Psi_q(y)}{q^2},
\end{equation}
with $\Psi_q(x) = 2\sin(q_1x_1)\sin(q_2x_2) $ 
and $q_i=k_i\pi/L$, $(k_i\in$ {\bf N}). Plugging (\ref{green})
back into (\ref{class_act_src}), we can calculate the value of the
action and the potential
energy $E^{(cont)}=2  {\mathcal{A}}\simeq -0.035144 \cdot F^2$. For
$\kappa\rightarrow\infty$, it is easy to check that the perfect action 
(\ref{perf_lapl_conf_space}) reproduces this continuum result exactly even
for $N=1$, i.e.~when the lattice consists of only one 
single point!

To compare classical quantities like the value of the action on
the lattice and in the continuum, we have to take into account the constant
$c$ which appears in the RG transformation (\ref{generalrgt}). For the fine
lattice action 
\begin{equation}
{\mathcal{A}}(\phi)=\frac{1}{2}\sum_{n,r}\rho^*(r)\phi_n\phi_{n+r}+f_\phi\sum_n
\phi_n, 
\end{equation}
with the perfect Laplace operator $\rho^*(r)$ and constant mass density $f_\phi$, we get after one RGT step the coarse
lattice action 
\begin{equation}
{\mathcal{A}}(\chi) = \frac{1}{2}\sum_{n_B,r_B}\rho^*(r_B)
  \chi_{n_B}\chi_{n_B+r_B} 
  + f_\chi\sum_{n_B}\chi_{n_B} - \sum_{n_B}\frac{f_\chi^2}{8\kappa},
\end{equation}
with $f_\chi=4f_\phi$. Hence, after iterating to the continuum, the
potential energy is calculated from the lattice field configuration $\bar\phi$
which solves the perfect Poisson equation by
\begin{equation} \label{potential_energy}
E = f_\phi\sum_n \bar\phi_n + \frac{F^2}{3 V\kappa},
\end{equation}
where $V$ denotes the lattice volume.

\begin{figure}[htb]
\begin{center}
\psfrag{yaxis}[][]{\hspace{-1.4cm}$|\frac{E-E_c}{E_c}|$}
\psfrag{1/n2}{\vspace{-0.3cm}$1/N^2$}
\psfrag{0.000}{\hspace{3mm}\footnotesize $0$}
\psfrag{0.005}{\hspace{-3mm}\footnotesize \ $0.005$}
\psfrag{0.010}{\hspace{-3mm}\footnotesize \ $0.010$}
\psfrag{0.015}{\hspace{-3mm}\footnotesize \ $0.015$}
\psfrag{0.020}{\hspace{-3mm}\footnotesize \ $0.020$}
\psfrag{0.00}{\hspace{2mm}\vspace{-1mm}\footnotesize $0$}
\psfrag{0.01}{\hspace{-1mm}\vspace{-1mm}\footnotesize \ $0.01$}
\psfrag{0.02}{\hspace{-1mm}\vspace{-1mm}\footnotesize \ $0.02$}
\psfrag{Standard Laplacian}{\scriptsize Standard Laplacian}
\psfrag{Parametrized Perfect Laplacian12345678}{\scriptsize Param.~Perfect
  Laplacian}  
\includegraphics[width=8cm]{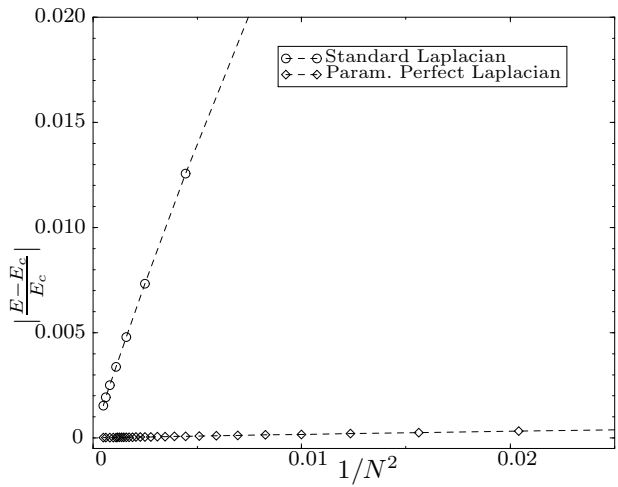}
\end{center}
\caption{Relative error for the numerical solution of our
  test problem. The error for the parametrized perfect
  Laplacian is for any resolution about 180 times smaller than for the
  standard Laplacian and gives a 0.06\% error at $N=5$. In other words, our
  parametrization gives better results on a $5^2$ lattice than the standard
  Laplacian on a $74^2$ lattice.} 
\label{plot_testres1}
\end{figure}

\begin{figure}[htb]
\begin{center}
\psfrag{yaxis}[][]{\hspace{-1.5cm}$|\frac{E-E_c}{E_c}|$}
\psfrag{1/n2}{\vspace{-0.4cm}$1/N^2$}
\psfrag{0.000}{\hspace{2mm}\vspace{-1mm}\footnotesize $0$}
\psfrag{0.002}{\hspace{-1mm}\vspace{-1mm}\footnotesize $0.002$}
\psfrag{0.004}{\hspace{-1mm}\vspace{-1mm}\footnotesize $0.004$}
\psfrag{0.006}{\hspace{-1mm}\vspace{-1mm}\footnotesize $0.006$}
\psfrag{0.008}{\hspace{-1mm}\vspace{-1mm}\footnotesize $0.008$}
\psfrag{0.010}{\hspace{-1mm}\vspace{-1mm}\footnotesize $0.010$}
\psfrag{0.012}{\hspace{-1mm}\vspace{-1mm}\footnotesize $0.012$}
\psfrag{0.00000}{\hspace{1mm}\footnotesize $\enspace\enspace0$}
\psfrag{0.00005}{\hspace{-6.5mm}\footnotesize $\enspace\enspace5\cdot 10^{-5}$}
\psfrag{0.00010}{\hspace{-8mm}\footnotesize $\enspace\enspace10\cdot 10^{-5}$}
\psfrag{0.00015}{\hspace{-8mm}\footnotesize $\enspace\enspace15\cdot 10^{-5}$}
\psfrag{S0 and S4 corrected for all points}{\scriptsize Normalization in all points}
\psfrag{S0 and S4 corrected for inner points only123456789}{\scriptsize
  Normalized only inner points} 
\includegraphics[width=8cm]{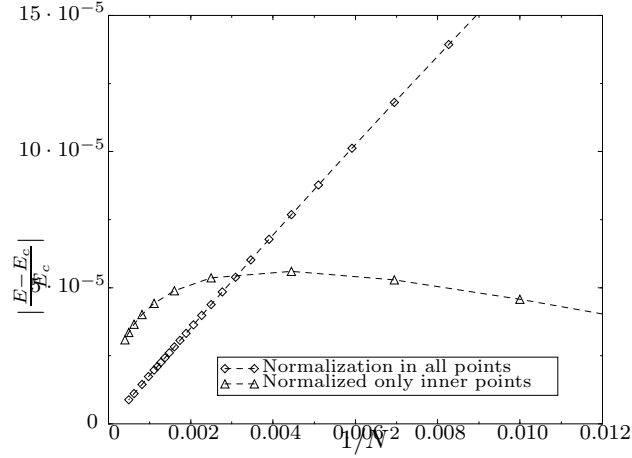}
\end{center}
\caption{Relative errors for the correctly normalized parametrized perfect 
  Laplacian and for an
  incompletely normalized parametrization. Although the correct
  normalization gives somewhat larger errors for
  small lattices (large lattice unit $a$), it makes sure that 
  for any lattice size, the results are better than for the standard
  Laplacian. The error of the standard Laplacian is far beyond the scale of
  this plot (compare Figure \ref{plot_testres1}).} 
\label{plot_testres2}
\end{figure}

As a check of the quality of our parametrization of the
perfect Laplace operator,
we numerically solved the lattice Poisson equation $A\phi=f$ for square
boundaries with the standard
$(4,-1,-1,-1)$-Laplacian and with our parametrized Laplace operator made
up from $\rho_1,\dots,\rho_{6}$. For the resulting field configuration, we
determined the potential energy $E= f\sum_n \bar\phi^{(std)}_n$ 
for the field configuration $\bar\phi^{(std)}$ computed with the standard
Laplacian and (\ref{potential_energy})
for the configuration $\bar\phi^{(pp)}$ computed with the parametrized perfect
Laplacian. 
Fig.~\ref{plot_testres1} shows a plot of the relative error
$(E-E^{(cont)})/E^{(cont)}$ as a 
function of the inverse lattice volume $1/N^2$. The error for our
parametrization is proportional to $1/N^2$ and is for any lattice
size smaller by a
factor of about 180 than the error of the standard Laplacian.
This factor can probably still be increased by tuning the normalization
procedure of the truncated couplings.
Fig.~\ref{plot_testres2} shows what happens when the couplings are not
normalized and therefore do not 
fulfill the sum rules $S_0$ and $S_2$. We see that the error is then no longer
proportional to the inverse lattice
size. While for coarse lattices --- that is small $N$ --- the error is small, the results get worse for
finer lattices, and somewhere a point is reached where the standard Laplacian
gets better than the parametrization. If this should be avoided, one has to
correct the truncation error.

\begin{figure}[htb]
\begin{center}
\psfrag{E}[][]{\hspace{-5mm}$E$}
\psfrag{1/N}{\vspace{-3mm}$1/N$}
\psfrag{-0.022}{\hspace{-2mm}\scriptsize -0.022}
\psfrag{-0.024}{\hspace{-2mm}\scriptsize -0.024}
\psfrag{-0.026}{\hspace{-2mm}\scriptsize -0.026}
\psfrag{-0.028}{\hspace{-2mm}\scriptsize -0.028}
\psfrag{-0.030}{\hspace{-2mm}\scriptsize -0.030}
\psfrag{-0.032}{\hspace{-2mm}\scriptsize -0.032}
\psfrag{0.00}{\vspace{-1mm}\scriptsize \ \ 0}
\psfrag{0.05}{\vspace{-1mm}\scriptsize 0.05}
\psfrag{0.10}{\vspace{-1mm}\scriptsize 0.10}
\psfrag{0.15}{\vspace{-1mm}\scriptsize 0.15}
\psfrag{Standard Laplacian}{\scriptsize Standard Laplacian}
\psfrag{Parametrized Perfect Laplacian12345678}{\scriptsize Param.~Perfect Laplacian} 
\includegraphics[width=8cm]{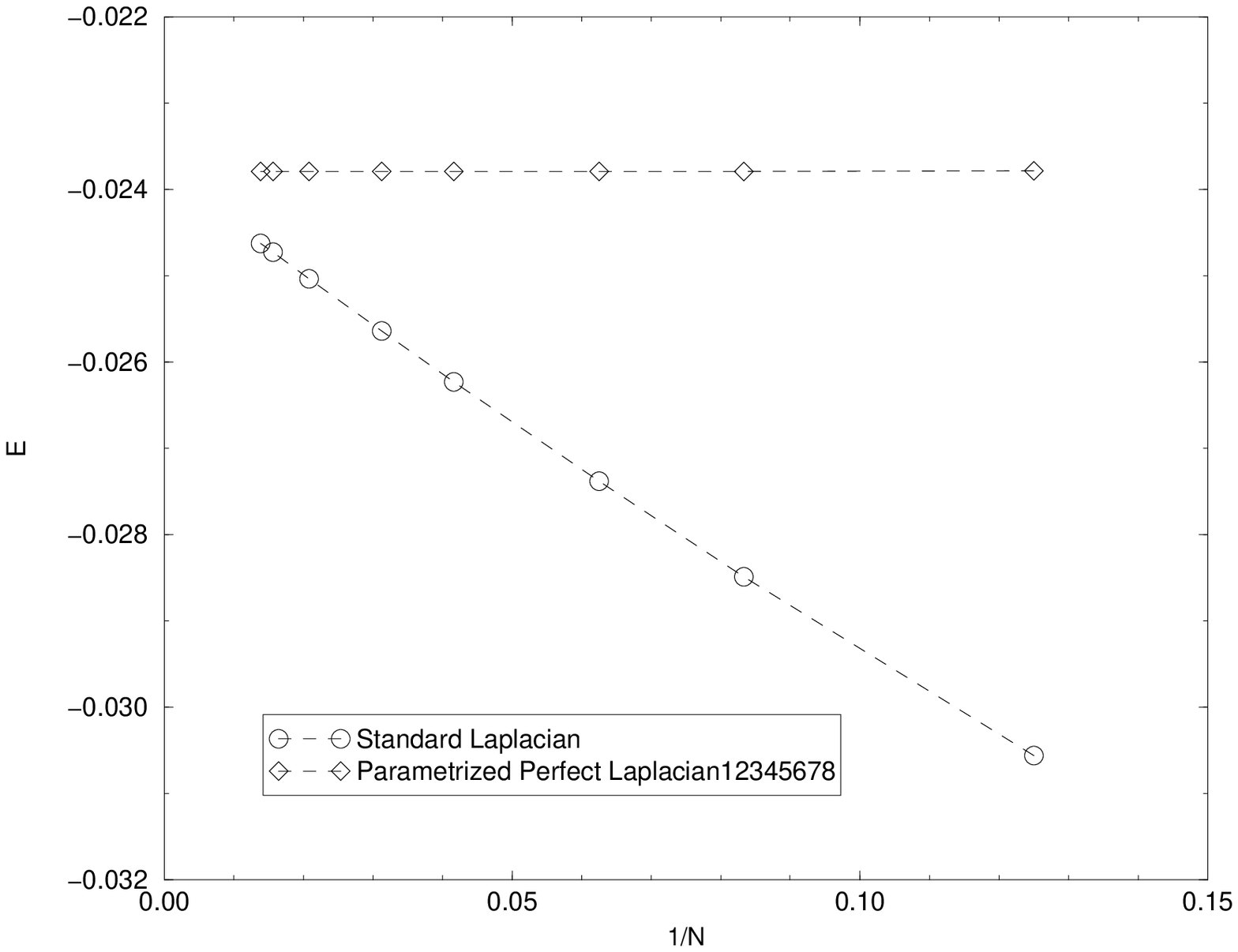}
\end{center}
\caption{Comparison of the results for the potential energy of an L-shaped
  membrane between standard and parametrized perfect
  Laplacian. The error for the standard Laplacian is of order $1/N$.}
  \label{fig_res_lshape} 
\end{figure}

Finally, we make the same comparison for 
L-shaped boundaries as shown in Fig.~\ref{fig_cclat}. As above, we solve the
lattice Poisson equation with the 
standard Laplacian and the parametrized perfect Laplacian, for which we take
the couplings $\rho_1,\dots,\rho_{13}$ listed in Tables \ref{table_cpl1} and \ref{table_cpl2}. 
The lattice volume $V$ for this shape is $V= 3/4 N^2$. 
We compare lattice sizes $N$ between 8 and 72.
As we see in
Fig.~\ref{fig_res_lshape}, the parametrized perfect Laplacian gives excellent
results for any lattice size, while for the standard Laplacian we have to use
very fine lattices to get acceptable results. For example, take a lattice with
$N=10$. The standard Laplacian then has an ${\mathcal{O}}(25\%)$ error, while the error
for the parametrized perfect Laplacian is ${\mathcal{O}}(0.1\%)$. 

\vspace*{1pt}\textlineskip 
\section{Generalizations} 
\vspace*{-0.5pt}
\noindent
The parametrization presented in the previous sections has some
limitations: The boundaries have fixed orientation and are always half a
lattice spacing away from 
the lattice points, and we only used a constant source in the action. In this
section we show how to overcome these limitations.

Consider a RG transformation of the action with a general source term $J(x)$:
\begin{equation}
   S_J[\phi] = \frac{1}{2}\int d^dx\ [\partial_\mu\phi(x)\partial_\mu\phi(x) +
   J(x)\phi(x) ].
\end{equation}
Blocking out of continuum
\cite{wilson:76,bietenholz_wiese:96,bietenholz_wiese:96-2} with the kernel 
\begin{equation}
T_\kappa[\Phi,\phi] =
\kappa\sum_n\Bigl(\Phi_n - \int d^dx\ \omega(x-n)\phi(x)\Bigr)^2,
\end{equation}
where $\omega(x)$ is an arbitrary blocking function gives the the result in
Fourier space
\begin{equation} \label{eq_reslt2_exactrgt}
  {\mathcal{A}}[\Phi] = 
  \frac{1}{
    2}\int\limits_{-\pi}^{\pi} \frac{d^dk}{(2\pi)^d} \tilde\Phi(k)
    \tilde\rho^*(k) \tilde\Phi(-k) -
    \int\limits_{-\infty}^{\infty}\frac{d^dk}{(2\pi)^d} \tilde\Phi(-k)
    \tilde\rho^*(k) \frac{\tilde\omega(k)}{k^2} \tilde
    J(k)  - W(J^2).
\end{equation}
$W(J^2)$ denotes a $\Phi$-independent term where the source only appears in
second 
order and the inverse of $\rho^*(k)$ is the fixed point propagator
\begin{equation}
  \frac{1}{\tilde\rho^*(k)} = \sum_{l\in{\mathbf Z^d}}
  \frac{\tilde\omega(k+2\pi l)\tilde\omega(-k-2\pi l)}{(k+2\pi l)^2} +
  \frac{1}{\kappa}.
\end{equation}
The equation of motion to the fixed point action (\ref{eq_reslt2_exactrgt}) in
configuration space is
the perfect Poisson 
equation
\begin{equation}
  \sum_{n^\prime} \rho^*(n-n^\prime) \Phi_{n^\prime} = -J_n^{FP}.
\end{equation}
with 
\begin{equation}
 \tilde J^{FP}(k) = \tilde\rho^*(k) \frac{\tilde\omega(k)}{k^2} \tilde J(k),
\end{equation}
Therefore the perfect Laplace operator can be used to solve
the Poisson equation with any source $J(x)$.

From Eq.~(\ref{eq_reslt2_exactrgt}) we can also read the fixed point field
operator $\phi^{FP}$ which remains unchanged under RG transformations.
Assume that the source $J(x)$ is small. Then the term of order $J^2$ can be
neglected, and the transformation back to configuration space yields
\begin{equation}
  {\mathcal{A}}[\Phi]\ = 
  \frac{1}{2} 
  \int\limits_{-\pi}^{\pi}\frac{d^dk}{(2\pi)^d}
  \tilde\Phi(k)\tilde\rho^*(k)\tilde\Phi(-k) - \int
  d^dx J(x)\phi^{FP}(x) .
\end{equation}
The fixed point field operator is 
\begin{equation} \label{eq_contlatfd}
  \phi^{FP}(x) = \sum_n Z(x-n)\Phi_n,
\end{equation}
with the coefficient function $Z(x)$ in configuration space
\begin{equation} \label{eq_fpfield_coeff}
  Z(x) = \int\limits_{-\infty}^\infty \frac{d^dk}{(2\pi)^d} e^{-ikx}
  \tilde\rho^*(k) 
  \frac{\tilde\omega(k)}{k^2}.
\end{equation}

Using the fixed point field, we 
calculate the fixed point Laplacian in $d$ dimensions for an arbitrarily placed
$d-1$-dimensional hyperplane as a boundary.
Consider zero boundary conditions $\phi(x)=0$, $ \forall x\in\Gamma$, where
the boundary $\Gamma$ divides the hyperspace into 
the halfspaces $A$ inside and $B$ outside the
boundary (see Fig.~\ref{fig_planebd} for $d=2$).
\setlength{\unitlength}{0.8cm}
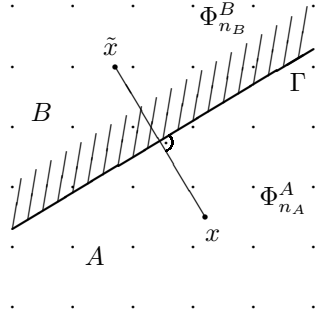
\begin{figure}[htb]
\begin{center}
\begin{picture}(5,5.2)
\multiput(0,0)(1,0){6}{
  \multiput(0,0)(0,1){6}{\circle*{0.05}}}
\thicklines
\put(0,1.3){\line(5,3){5}}
\put(4.6,3.6){$\Gamma$}
\thinlines
\multiput(0,1.3)(0.25,0.15){20}{\line(1,6){0.14}}
\put(1.2,0.7){$A$}
\put(0.3,3.1){$B$}
\put(3.2,1.5){\circle*{0.08}}
\put(3.2,1.1){$x$}
\put(3.2,1.5){\line(-3,5){1.5}}
\put(1.7,4){\circle*{0.08}}
\put(1.5,4.2){$\tilde x$}
\put(2.55,2.735){\circle*{0.05}}
\put(2.54,2.6){\qbezier(0,0)(0.18,0.1)(0.05,0.26)}
\put(4.1,1.7){$\Phi^A_{n_A}$}
\put(3.1,4.7){$\Phi^B_{n_B}$}
\end{picture}
\end{center}
\caption{The line $\Gamma$ as a boundary in $d=2$. In the text we
  show how to calculate the 
  perfect Laplacian restricted to the halfplane $A$.} \label{fig_planebd}
\end{figure}
Start with the identity 
\begin{equation} \label{ident}
  \int d^dx \left(\phi(x)+\phi(\tilde x)\right)^2 = 0,
\end{equation}
which is trivially fulfilled because we impose the condition $\phi(\tilde x)
= -\phi(x)$ on the field, where $\tilde x$ denotes $x$ mirrored at the
boundary. If we plug the fixed point field (\ref{eq_contlatfd}),
(\ref{eq_fpfield_coeff}) into Eq.~(\ref{ident}), we get the quadratic form
$\Phi^T C\Phi = 0$,
where $\Phi$ is the lattice field vector
and the elements of the symmetric matrix $C$ are
\begin{equation}
  C_{nm} = \int d^dx \left[Z(x-n)+Z(\tilde x-n)\right]\left[Z(x-m)+Z(\tilde x-
  m)\right].
\end{equation}
As it doesn't matter how we order the lattice points when arranging them in a
vector, the lattice field 
vector $\Phi$ can be split into the subvectors $\Phi^A$ that collects all
the 
components $\Phi_{n_A}$ living on the lattice 
points in halfspace $A$,
and $\Phi^B$ that collects the components $\Phi_{n_B}$ in
$B$. The quadratic form then gets
\begin{equation} \label{eq_phia_phib}
  (\Phi^A)^T C_A \Phi^A + 2(\Phi^B)^T C_{BA}\Phi^A + (\Phi^B)^T C_B \Phi^B =0, 
\end{equation}
where $C_A$, $C_{BA}$ and $C_B$ are submatrices of $C$
\begin{equation}
  C = \left( \begin{array}{cc} C_A & C_{AB} \\
      C_{BA} & C_B \end{array} \right),
\end{equation}
with $C_{BA}=C_{AB}^T$. Keeping $\Phi^A$ fixed and minimizing
Eq.~(\ref{eq_phia_phib}) in $\Phi^B$, we 
find the relation between lattice points outside and inside the boundary
$\Phi^B = -C_B^{-1}C_{BA}\Phi^A$. Plugging this into the perfect lattice
Laplace equation 
$  \sum_{n^\prime} \rho^*(n-n^\prime)\Phi_{n^\prime} = 0$, 
the sum over all lattice points $n^\prime$ is replaced by a sum over the
halfspace $A$
\begin{equation}
  \sum_{n^\prime_A} \left[ \rho^*(n-n^\prime_A) -
   \sum_{n^\prime_B} \rho^*(n-n^\prime_B)
  (C_B^{-1}C_{BA})_{n^\prime_B,n^\prime_A} \right] \Phi_{n^\prime_A} = 0,
\end{equation}
and
the perfect Laplacian $\rho_A^*(n_A,n^\prime_A)$ restricted to the halfspace
$A$ is given by the term in square brackets.
As the fixed point field operator is local, this exact result can be used to
construct another parametrization of the fixed point Laplacian. The position
and orientation of the hyperplane relative to the lattice are then the
parameters and it only remains to calculate the couplings numerically.

\vspace*{1pt}\textlineskip 
\section{Conclusion}
\vspace*{-0.5pt}
\noindent
The fixed point action has its applications not only in field theory, it 
can also serve as a powerful tool for the solution of partial differential
equations, where we encounter non-trivial boundary conditions. We showed that
the influence of boundaries on the fixed point action is highly
local. Therefore we could provide a parametrization for the perfect lattice
Laplace operator in $d=2$ which is easy to use and gives nearly perfect
results for any resolution. A generalization to $d$ dimensions is trivial, as
none of the calculations is specific to two dimensions. 
While the parametrization provided in this work can be used to
approximate boundaries of any form, it is not the only one possible, and we
proposed an alternative to find a parametrized fixed point Laplacian for
non-trivial boundaries. 
The benefit from using perfect discretizations is enormous, as we pointed
out for test problems.

\nonumsection{Acknowledgments}
\noindent
I thank Peter Hasenfratz and Ferenc Niedermayer for their ideas and support
during this work and Urs Wenger for his advice.

\nonumsection{References}

\end{document}